# Population stratification enables modeling effects of reopening policies on mortality and hospitalization rates
## - A case study in Houston, Texas, USA

Tongtong Huang[1], Yan Chu[1,1], Shayan Shams[1,1], Yejin Kim[1], Genevera Allen[2], Ananth V Annapragada[3], Devika Subramanian[2], Ioannis Kakadiaris[4], Assaf Gottlieb[1,+], Xiaoqian Jiang[1,+]

[1] School of Biomedical Informatics, UTHealth, Houston, TX
[2] Department of Computer Science & Electrical and Computer Engineering, Rice University, Houston, TX
[3] Department of Pediatric Radiology, Texas Children's hospital, Houston, TX
[4] Department of Computer Science, Electrical & Computer Engineering, and Biomedical Engineering
University of Houston, Houston, TX

## Abstract

**Objective:** We study the influence of local reopening policies on the composition of the infectious population and their impact on future hospitalization and mortality rates.

**Materials and Methods:** We collected datasets of daily reported hospitalization and cumulative morality of COVID-19 in Houston, Texas, from May 1, 2020 until June 29, 2020. These datasets are from multiple sources (USA FACTS, Southeast Texas Regional Advisory Council COVID-19 report, TMC daily news, and New York Times county-level mortality reporting). Our model, risk-stratified SIR-HCD uses separate variables to model the dynamics of local-contact (e.g., work from home) and high-contact (e.g., work on site) subpopulations while sharing parameters to control their respective $R_0(t)$ over time.

**Results:** We evaluated our model's forecasting performance in Harris County, TX (the most populated county in the Greater Houston area) during the Phase-I and Phase-II reopening. Not only did our model outperform other competing models, it also supports counterfactual analysis to simulate the impact of future policies in a local setting, which is unique among existing approaches.

**Discussion:** Local mortality and hospitalization are significantly impacted by quarantine and reopening policies. No existing model has directly accounted for the effect of these policies on local trends in infections, hospitalizations, and deaths in an explicit and explainable manner. Our work is an attempt to close this important technical gap to support decision making.

**Conclusion:** Despite several limitations, we think it is a timely effort to rethink about how to best model the dynamics of pandemics under the influence of reopening policies.

**GitHub Link:** https://github.com/th8930/ssirhcd
**R Shiny Weblink:** https://yan-chu.shinyapps.io/ssir-hcd_shiny/

---

[1] equal contribution to first authorship

[+] Joint corresponding authorship

# Introduction

COVID-19 has taken the international community by surprise[1]. At the time of writing this paper, the COVID-19 pandemic has surpassed 10 million confirmed cases and 500,000 deaths worldwide[2]. COVID-19 is having a dramatic impact on health care systems in even the most developed countries[3]. Without effective vaccines and treatments in sight, the only effective actions include policies of containment, mitigation, and suppression[4].

The infection, hospitalization, and mortality trends of COVID-19 across different countries vary considerably and are affected mainly by policy-making and resource mobilization[5]. Predicting the local trends of the epidemic is critical for the timely adjustment of medical resources and for the evaluation of policy changes in an attempt to curtail the economic impact[6]. In the United States, policies vary by state and city, and therefore, robust *local* models are essential for learning fine-grained changes that meet the needs of local communities and policymakers.

Under appropriate intervention, early studies observe a trajectory of consumption recovery near the end of the eight-week post-outbreak period (following the classical epidemiology models)[7]. However, traditional models do not account for the impact of local policies, such as a multi-phase reopening. The recent rebounds in Texas indicate different trends in different counties, which motivated the need to study the underlying impact of policy on local mortality and hospitalization trends. In this paper, we present the design of our regional model and demonstrate its use by applying them to the Houston, TX area marking their difference from the global trend estimation models.

Due to the lack of consistent and accurate estimations of infection rates in asymptomatic individuals (using, e.g., random serological testing[8]), we are focusing on mortality and hospitalization. We present the development of a forecasting model using local fine-grained hospital-level data to track the changes in hospitalization and mortality rates owing to reopening orders in the greater Houston area encompassing nine counties in the state of Texas, USA. The modeled area consists of 4,600 km$^2$, incorporating a population of 3,012,050 adults and 1,080,409 children (by the 2010 census) and includes over 100 hospitals with a total bed capacity of 23,940[9,10]. Our methodological contribution is directly modeling the impact of phased reopening. We achieve it by splitting the targeted population into low-contact and high-contact groups (determined by the subpopulations that return to work at different phases of the reopening). The mechanism adjusts the proportion of infectious subpopulations (depending on their category of jobs) to quantitatively represent the policy impact on the epidemiological dynamical system (please refer to Figure 1 for a high-level overview). It can built into most existing epidemiological models without ease, offering additional explanation ability and better prediction efficacy. We demonstrated our new approach using a policy-aware risk-Stratified Susceptible-Infectious-Recovered Hospitalization-Critical-Dead (SSIR-HCD) model, which compared favorably to existing methods (including our neural network latent space modeling, a nonlinear extension of SIR-HCD).

## Related work

There are many predictive models for COVID-19 trend prediction. The Center for Disease Control (CDC) also hosts 23 different trend predictors[11,12] to forecast total death. There are several big categories:

- Purely data-driven models (with no modeling of disease dynamics), which includes regression-based parametric and non-parametric models (Auto-Regressive Integrated Moving Average or ARIMA, Support Vector Regression, Random Forest), neural network (deep learning) based trend prediction (e.g., GT-DeepCOVID[13]), etc.
- Epidemiology based dynamic models based on grouping populations into a discrete set of compartments (i.e., states), and defining ordinary differential equations (ODE) rate equations describing the movement of people between compartments: SEIR (Susceptible, Exposed, Infected, Recovered) models and their myriad variants are examples in this category.
- Individual-level network-based models: finest grain modeling of a population through agent simulation, such as the ones built in NetLogo by Marathe *et al.* [14] and NotreDame-FRED[15].
- Various ensemble and hybrid models: including the Imperial College London short-term ensemble forecaster[11] and IHME model [16] that combines a mechanistic disease transmission model and a curve-fitting approach.

Among existing models, the ODE compartment-based models occupy a middle ground between network models at the individual-level and purely count-driven statistical analyses that are disease-dynamics-agnostic, which will be our main interest in this paper. Compartment models, which originated in the early 20th century[17], still represent the mainstream in epidemiological studies of infectious disease. They make a critical mathematical simplification by decomposing the entire population into compartments (i.e., states), e.g., susceptible, infectious, recovered, and use ODEs to model the transitions between the compartments (Table 1). These compartment models make assumptions that the observation counts in the various compartments naturally reflect the reproduction number $R_0$ that changes over time. The recent COVID-19 pandemic, however, has introduced the need to incorporate lockdown policy interventions (i.e., how long the population will remain at home), which existing compartment models have not considered. We observe different patterns of hospitalization and mortality even within a single metropolitan area such as Houston, TX, which means traditional epidemiology systems might not be sufficient to explain the dynamics. Many people speculated that local policies (shutdown and reopening) could have introduced perturbations to the disease dynamics. Still, it is not clear how to quantify their impacts and provide counterfactual reasoning to support future policy decisions. Our SSIR-HCD is a unique effort to close the modeling gap by using appropriate data to enrich the established compartment models. The only other relevant model [18] focused on anti-contagion policies, which is significantly different from our phased reopening policies model in that we considered the stratified risks in the population (related to people who might have more chances of exposure, depending on the phases in the reopening policy).

## Data and Materials

We collected experimental datasets of the daily reported hospitalization and cumulative morality of COVID-19 that occurred in Houston, Texas, from May 1, 2020 (the start date of Phase 1 reopening in Houston, TX) until June 29, 2020. Population data was collected by USA FACTS[23], industry employment data was gathered from U.S. BUREAU OF LABOR STATISTICS[24], and the hospitalization data sources originate from Southeast Texas Regional Advisory Council (SETRAC) COVID-19 report[25]. We used TMC daily news[26] to set the initial length of hospitalization for our model. We also used mortality data from The New York Times county-level report[27]). Note that New York Times data combine confirmed and suspected cases in their reporting of mortality. To be consistent, we used SETRAC hospitalization reporting that contains both confirmed and suspected cases.

In this study, we focused on the data from Harris County, one of the nine counties in Houston, TX with the largest population.

## Method

We propose a forecast model based on SIR-HCD with a novel variant on compartments to address the differences in local policy. In SIR-HCD, the entire population is divided into six sub-groups: susceptible population $S$, exposed population $E$, infectious population $I$, recovered population $R$, hospitalized population $H$, critical population $C$, and dead population $D$. The transitions between sub-groups are governed by nonlinear ordinary differential equations. Please refer to Table 2 for our nomenclature.

### SIR-HCD overview

We use the SIR-HCD to model the state transitions. The model is a simplification of SEIR-HCD. We decide to drop exposed state (E), which cannot be reliably modeled in COVID-19 because the CDC guideline for exposure, determined as staying within less than six feet for more than fifteen minutes from a person with known or suspected COVID-19[11], is too short a time period to be modeled adequately. Thus, a simpler SIR-HCD model, which assumes the possibility of direct transitions between the susceptible state and the infectious state, is more suitable in COVID-19.

In the SIR-HCD model, some susceptible people may turn into an infectious status after the incubation period. Infectious people may either get hospitalized or recover after a certain period of time. A proportion of the hospitalized people might be admitted to the Intensive Care Unit (ICU), while the rest of them will recover in the hospital. Similarly, among the critical cases (i.e., ICU patients), some people might die, and others will recover. Thus, the SIR-HCD model follows a series of nonlinear ODEs to model the state transitions:

$$\frac{dS(t)}{dt} = -\frac{R_0}{T_{inc}} I(t) S(t),$$

$$\frac{dI(t)}{dt} = \frac{R_0}{T_{inc}}I(t)S(t) - \frac{I(t)}{T_{inc}},$$

$$\frac{dH(t)}{dt} = (1-r_a)\frac{I(t)}{T_{inc}} + (1-r_f)\frac{C(t)}{T_{crit}} - \frac{H(t)}{T_{hosp}},$$

$$\frac{dC(t)}{dt} = \frac{r_c H(t)}{T_{hosp}} - \frac{C(t)}{T_{crit}},$$

$$\frac{dR(t)}{dt} = \frac{r_a I(t)}{T_{inc}} + (1-r_c)\frac{H(t)}{T_{hosp}}, \text{ and}$$

$$\frac{dD(t)}{dt} = \frac{r_f C(t)}{T_{crit}}.$$

Note that $R_0(t)$, which is shorted as $R_0$ and used interchangeably in our paper, denotes a dynamically changing reproduction number (considering several changes of quarantine policies published in Houston). The symbol $T_{inc}$ denotes the average incubation period of COVID-19. In the equations that models $H, C, R, D$, the term $T_{hosp}$ represents the average time that a patient is in a hospital before either recovering or becoming critical, and $T_{crit}$ denotes the average time that a patient is in a critical state before either recovering or dying. In addition, $r_a$ refers to the asymptomatic rate in infected populations $I$, $r_c$ refers to the critical rate in hospitalized population H, and $r_f$ refers to the deceased rate in critical population $C$. This model is more robust than SIR, as the introduction of more reliable observations of $H, C, D$ provides extra stabilization to the dynamic system. Figure 2 illustrates the SIR-HCD model with its basic states and transitions implied by the ODE function.

With reopening policies in place, there are more interactions between people and so the likelihood of spread increases. Our expectation is either that $R_0$ remains constant (because people maintain safe distances and follow CDC protocols), or (more likely) that it increases with spotty compliance with pandemic protocols. To make the computation tractable, we decide to use the inverse operation of the exponential Hill decay equation to model $R_0$ as following,

$$R_0(t) = (1 + (t/L)^k) \cdot R_0(0)$$

where $L$ refers to the rate of decay, and $k$ controls the shape of the decay. When $k = 1$, the above exponential equation is just a monotonically increasing linear equation. We set the starting point $t = 0$ as the reopening date, May 1, 2020. The initial states $H(t = 0)$ and $D(t = 0)$ are the numbers of reported hospitalized cases and cumulative mortality in Harris County on that date.

We decided not to rely on confirmed cases, assuming that the actual number for the infected population is larger than the reported number (such an effect has been reported in California[28] and New York[29]). Since a fraction of the actual infected patients were hospitalized on the first day, the initial infectious population $I(t = 0)$ is therefore estimated to be $m$ times the initially hospitalized number $H(t = 0)$, where $m$ is a positive constant coefficient. Some studies suggested that true positive infectious cases should be 50 - 90 times more than the reported positives [30,31]. In the Harris County projection, we set $m$ to be 60, assuming that $H(t = 0)$ is approximately equal to "known positives" on the first day. To estimate the recovery rate, we divided Harris' case mortality rate (the number of confirmed deaths on the current day) by the number of confirmed cases 14 days before that, as reported by The New York Times[27]. The average mortality rate starting from May 1, 2020, was 2%. Therefore, we have an estimated recovery rate of 98%. In this case, the initially recovered individuals $R(t = 0) = 0.98 \cdot I(t = -14) = 0.98m \cdot H(t = -14)$, where $t = -14$ refers to 14 days earlier than the starting date (i.e., April 17, 2020). The number of critical individuals $C(t = 0)$ is set to be 50% of hospitalized individuals $H(t = 0)$ based on the average proportion of ICU usages among COVID-19 hospitalization in Texas[11,25]. The initial number of susceptible population $S(t = 0)$ is

$$S(t = 0) = N - I(t = 0) - R(t = 0) - H(t = 0) - C(t = 0) - D(t = 0)$$

where $N$ is the total population in the county.

Following a previous SIR-HCD optimization method[22], we used the limited memory Broyden–Fletcher–Goldfarb–Shanno algorithm (L-BFGS-B)[32] to optimize the ODE system. According to previous COVID-19 studies[33], the constant parameter $T_{inc}$ is set to 14 days. The optimal values of parameters $R_0, T_{hosp}, T_{crit}, r_a, r_c, r_f, L$, and $k$ in the model were obtained by minimizing the weighted average mean squared log error (MSLE) loss function $L(MSLE)$. To make the prediction more focused on the recent trajectory, we used the squared log error at each time point with a weight parameter satisfying the condition $W_t > W_{t-1}$. Finally, we used a time inverse function $W_t = 1/(t_{max} - t + 1)$ in our model, where $t_{max}$ is the maximum time.

$$L(MSLE) = 1/T \sum_{t=0}^{T} W_t((log(H(t)) + 1) - (log(H'(t)) + 1))^2 + 1/T \sum_{t=0}^{T} W_t((log(D(t)) + 1) - (log(D'(t)) + 1))^2$$

## SSIR-HCD model to explicitly account for local policy's impact

In this section, we introduce the unique aspect of our model that differentiates it from existing ones. Our intuition here is that people get infected either through family transmission or through social (including job) activities. In the transition from a strict stay-at-home to reopening, the population is subject to changes in their social activities, which impact their probability of infection as well as their risk of transmission to their family members. Therefore, we can divide the total population in Harris County into two groups; a low-contact group, which includes people in industries that were still closed (e.g., working from home subpopulation and their families, including those who are unemployed but not homeless), and a high-contact group includes people in industries that were reopened due to economic restart (e.g., working on site subpopulation and their families). Intuitively, the subpopulation of people who work from home is those who continue to stay at home and have limited chances of contacting the working subpopulation.

The two groups share the same fitted parameters $R_0, T_{hosp}, T_{crit}, r_m, r_c, r_f$, as well as the same constant incubation period $T_{inc}$, but they are estimating different $R_0$. We set the initial $R_0(t = $

0)for the low-contact group is slightly lower than that of the high-contact group, and low contact $L$ is slightly higher than or equal to the high-contact group. This is to keep the low-contact $R_0$ being differentiated from high-contact $R_0$ overtime. The unique coupling strategy makes it possible to directly reflect the impact of policy into SSIR-HCD (the superscript means squared as we model two subpopulations in the joint SSIR-HCD model).

According to reopening announcements released on the Texas government website[34] and the Houston employment rates by industry (reported by the Greater Houston Partnership Research[34,35]), necessary industries such as transportation, utilities, government, and a subset of the health services kept running before and during the reopening of the economy, accounting for 32.3% of the population in Houston. After releasing Reopening Phase I policies (May 1, 2020), 100% of the essential industries reopened, in addition to 15% health services, 25% professional and business service, and 25% leisure and hospitality, constituting a working on site (high-contact) subpopulation proportion of 39.62% after subtracting the unemployment rate of 0.4%[36]. The proportion of the high-contact population after Reopening Phase II (May 18, 2020) was a combination of 100% of the essential industries, 100% health services, 50% of professional and business service, and 50% of leisure and hospitality industries. Hence, the high-contact proportion among Reopening Phase II was 58.3% after subtracting the unemployment rate. Our model accounts for the change of low-contact and high-contact subpopulations between Reopening Phase I and Reopening Phase II, therefore directly modeling the policy's impact on epidemiological data over time.

## Experimental setting

Our training process uses MSLE to minimize the errors in curve-fitting. Additionally, we evaluated mean squared error (MSE), but it was not used for the curve-fitting process. As the training period is very short, and the observation data is highly volatile, we do not directly use the raw daily reported data for our training and forecasting. Similar to early work conducted by the School of Public Health at UTHealth [37], we also observed some data bumps (i.e., a large number of cases counted on one date instead of spread over time) in the reported hospitalization and mortality. Following the same consideration to avoid the influence of unreliable data on our modeling, we used a 7-day rolling average to smooth the raw inputs (and generating the training hospitalization and mortality data in the experiments).

Figure 3 shows the 7-day rolling average hospitalization and cumulative mortality from March 25, 2020, to June 29, 2020. The Texas government started phased-in reopening of the state on May 1, 2020 (the stay-at-home order was issued on March 31, 2020), then continued to expand reopening industries on May 18, 2020. Following these reopening phases in Texas, the daily hospitalization curve in Harris County was divided into three phases:

(1) Before Reopening Phase: March 25 - May 1, 2020
(2) Reopening Phase I: May 1 - May 17, 2020
(3) Reopening Phase II: May 18 - June 29, 2020

The hospitalization curve represents a delayed epidemic effect since the publication of the strict stay-at-home order on March 30, 2020. After reopening policies were issued in Texas (May 1, 2020), their impacts start to impact the dynamics in Reopening Phase I and Reopening Phase II.

Our local hospitalization and mortality modeling aims to fit the most recent phases (i.e., Reopening Phase I and Reopening Phase II) starting from May 1, 2020, to June 29, 2020. We validated the accuracy with data between June 23 and June 29. For comparison, our baselines were time-series regression models (exponential smoothing, autoregression, and ARIMA) and vanilla SIR-HCD. We predicted hospitalization and mortality, respectively, in the time-series regression models because they lack the capability to account for hospitalization and mortality together in one model. We also included our own Neural Network SIR-HCD model, which is equally flexible as SSIR-HCD. Interested readers can find the details in the Appendix.

## Results

Trained with Harris County cumulative hospitalization and mortality data in Reopening Phase I and Reopening Phase II, our SSIR-HCD model fits the trends in the training data well: Reopening Phase I (MSE=27.67 for hospitalization, MSE=1.57 for mortality) and Reopening Phase II (MSE= 5.20 for hospitalization, MSE=0.81 for mortality). As Figure 4 shows, the local hospitalization and mortality training curves are very close to the reported data, and the test curves also follow the data trends closely, which indicates our model is not overfitting to the training period.

Table 3 shows the prediction accuracy of the baseline models and risk-stratified SIR-HCD (SSIR-HCD) model. For the hospitalization prediction, the proposed SSIR-HCD model had a significantly higher accuracy (MSE=8.04) compared to the baselines (MSE=649.63, 22.48, 20.98 for three time-series regressions, and MSE=54.04 for vanilla SIR-HCD). For mortality prediction, we found that the time-series regression models generally predict well, and our proposed model had comparable accuracy. This high accuracy in mortality prediction of the general time-series models is mainly because the mortality rates were more stable than the hospitalization curve over time.

Table 4 displays the fitted values of eight training parameters in SSIR-HCD equations for the low-contact group and the high-contact group. These fitted parameter values correspond well to the values obtained in previous studies of COVID-19 [11],[38],[39]. And the ratio of hospitalizations turning into critical is close to the average ICU proportion among hospitalizations in Harris County, which was 50% in our initial state settings [11,25]. The constant parameter $T_{inc}$ is set at 11.5 in both groups based on the values suggested by the World Health Organization (WHO)[40] and the CDC[11]. As a sanity check, the $R_0$ values in the low-contact group are indeed lower than those values in the high-contact group, indicating a lower expected number of cases directly infected by individuals in the low-contact group.

## Counterfactual analysis

Figure 5 displays the SSIR-HCD model's counterfactual analysis results (of our model) on what would have happened in the absence of reopening policies after 160 days on May 1, 2020. In the x-axis, day 0 refers to May 1, 2020, day 17 refers to May 18, 2020, and day 60 refers to June 29, 2020. We restored the proportion of low-contact people and high-contact people to the no-reopening status (corresponding to 31.90% high-contact proportion of the population) while keeping all the trained parameters the same. Upon excluding all changes resulting from the reopening policies, it is noted that both modeled hospitalization and mortality curves become dramatically flat. The hospitalization curve with intervention reaches its peak on day 90, reducing nearly 2,500 existing cases. This demonstrates that quarantine policies are effective in controlling the spread of coronavirus as well as reducing the number of hospitalizations and mortality rates. Similarly, Figure 6 displays the counterfactual estimations on what would have happened if the Texas government did not continue to reduce limitations in Reopening Phase II. In Figure 6, the presumed reopening policies in Reopening Phase I represent moderate control to the hospitalization and mortality curves, reducing nearly 1,500 existing cases. Since a long stay-at-home order is not economically practical, our counterfactual analysis demonstrates that moderate reopening policies, keeping essential quarantine measures (such as mask order adoption), and opening several industries to lower capacity, may offer a reasonable middle ground between the strict quarantine and fully open economy. The chart of dynamic $R_0$ values show how dynamic $R_0$ differentiates the low-contact group and high-contact group such that modeled hospitalization and mortality curves would be flattened by increasing the proportion of the low-contact population. The model does not use one single reproduction number value to measure the integral transmission rate as the two subgroups have different levels of risks for getting infected.

## Limitation and Conclusion

Our SSIR-HCD model forecasts fine-grained COVID-19 hospitalization and mortality by accounting for the impact of local policies. One challenge is that the SSIR-HCD model is very sensitive to the initial values of $S$, $I$, $R$, $H$, $C$, and $D$ as the number of infectious agents is non-zero at the initial time point. We have managed to avoid overfitting the local time-series curve by deploying values based on the accumulated knowledge of these initial variables and also using a smoothed time series as a rolling 7-day average to alleviate the fluctuations. After variable adjustment, the predictive results obtained a low error rate, while also obtaining parameters that are close to real-world values, such as the asymptomatic rate $r_a$ that is close to 93.8% in the COVID-19 Scenarios outcome summary[25].

In publicly reported data, the cumulative mortality data in Reopening Phase II do not perfectly follow the hospitalization trends. Our expectation was that it would lag after the hospitalization cases by approximately 14 days. The actual mortality rate fluctuated in the middle of Reopening Phase II (despite we already smoothed the curve) when the number of hospitalization cases started to increase rapidly. Nonetheless, our SSIR-HCD model still approximates the hospitalization and mortality trends better than competing models. Thus, our model is advantageous over baseline regressions. It can fit epidemiological data with complicated shapes, such as Harris hospitalization data, based on the proportion of low-contact and high-contact groups and can consider several epidemiological states together into one model that can make predictions for one or more sub-populations simultaneously. In addition to forecasting, our model offers another unique functionality to support counterfactual analysis, which can be useful in supporting critical decision-making.

However, our SSIR-HCD model inherited SIR-HCD in assuming a monotonically increasing $R_0$. This assumption has limitations to the future when economic reopening might be paused due to the overestimation of $R_0$ (facing a big susceptible population). For example, if a local policy were to clamp down on exposure (e.g., mandating masks and other means to influence infectivity), it is not reflected in SSIR-HCD, which is an obvious weakness. One possible strategy is to introduce an adjustable $R_0$ control to the model, such as our extended model called Neural Network SIR-HCD (See Appendix), which learns the quarantine strength over time to determine $R_0$ change. Additionally, our model interprets the recovered population as those who can no longer infect other individuals under the condition that the number of susceptible individuals keeps decreasing over time. We did not consider the possibility that some COVID-19 survivors may be reinfected after they were recovered, which could influence the modeling coronavirus transmission rate. Several of these aspects involve controversial discussions in the scientific community, but a powerful model should be able to accommodate different assumptions.

There are other reality constraints that our model is not taking into consideration. For example, the number of daily hospitalizations and critical patients cannot increase without limit due to total bed capacity in hospitals. In fact, Texas Medical Center reported they reached 100% of ICU basis capacity on June 25, 2020 [41]. Our model did not consider hospitalization and ICU delays when some hospitals are fully loaded, which needs more model parameters.

Yet another limitation of our model is the lack of full consideration for population density, demographics composition, daily in-bound/out-bound traffic flows, and medical resource disparities. For example, many patients in Harris County might come from other counties, but they are treated in the Texas Medical Center (in Harris County), so the total hospitalization and mortality might not completely match the local infection rates. Joint consideration of multiple counties and decomposition of hospitalized patients in terms of their residency would produce more accurate predictions.

We have presented a proof-of-concept of a policy-aware compartmental dynamical epidemiological model by stratifying populations into low- and high- risk groups based on people's affiliated industries during the reopening phases at a county level using limited data. We believe it is an important effort to better understand dynamic feedback of this stratification through an ODE control system. There are many limitations and future directions that we have exposed through this exploration. We will further explore these challenges with more data and better assumptions to improve existing models.

# Appendix

**GitHub Link:** https://github.com/shayanshams66/NN-SIR-HCD [Neural Network SIR-HCD]

**Neural Network SIR-HCD model (with adjusted quarantine control)**

Since the controlling parameter $R_0$ within the SIR-HCD model does not account for specific quarantine effects. In reality, it is reasonable to consider the quarantine factors for adjusting free parameters in our existing SIR-HCD model. Therefore, we utilize a Multilayer Perceptron (MLP) architecture [39], [42] to estimate the hidden variable $Q$ and augment the epidemiological estimation process. The augmented model introduces a quarantine strength term $Q$ and quarantined population $T$. We designed $Q(t)$ as an n-layer MLP network with a weighted vector $W$, and the input vector $x(t) = (S(t), I(t), R(t), H(t), C(t), D(t), T(t))$. Therefore, the hidden variable $Q(t)$ is estimated as

$$Q(t) = NN(x(t), W)$$

The original reproduction number $R_0$ at each timestep $t$ is the constant value. We aim to adjust the value of $R_0$ by adding the variant quarantine strength term $Q(t)$ so that the curve could be more flexible to fluctuate policy changes.

$$R_0' = R_0 + Q(t) = R_0 + NN(x(t), W)$$

We utilized a Multilayer perceptron (MLP) network with two hidden layers in our implementation. The first and second hidden layers each have 10 and 20 neurons with RELU activation function (See Figure s1). The hidden variable $Q(t)$ is estimated using the Neural Network. The new ordinary differential equations for the augmented SIR-HCD model are

$$\frac{dS(t)}{dt} = -\frac{R_0 + Q(t)}{T_{inc}} I(t) S(t)$$

$$\frac{dI(t)}{dt} = \frac{R_0 + Q(t)}{T_{inc}} I(t) S(t) - \frac{I(t)}{T_{inc}}$$

$$\frac{dH(t)}{dt} = (1 - r_a)\frac{I(t)}{T_{inc}} + (1 - r_f)\frac{C(t)}{T_{crit}} - \frac{H(t)}{T_{hosp}}$$

$$\frac{dC(t)}{dt} = \frac{r_c H(t)}{T_{hosp}} - \frac{C(t)}{T_{crit}}$$

$$\frac{dR(t)}{dt} = \frac{r_a I(t)}{T_{inc}} + (1 - r_c)\frac{H(t)}{T_{hosp}}$$

$$\frac{dD(t)}{dt} = \frac{r_f C(t)}{T_{crit}}$$

$$\frac{dT(t)}{dt} = Q(t) I(t)$$

The starting point for the variables $S(t = 0), I(t = 0), R(t = 0), H(t = 0), C(t = 0), D(t = 0)$ is the same as ones in the idealized SIR-HCD model. The quarantined population is initialized to a small value $T(t = 0) = 10$, and the neural network model learns how to estimate it based on local data in each county.

The deep learning adjusted SIR-HCD model is trained by minimizing the weighted mean squared log error loss function using the ADAM optimizer [43] for 1000 iterations. The loss function calculates the weighted average squared error, in which the weight $W_t$ at a later time has a higher value. The optimization continues until the loss value converges.

$$L(MSLE) = 1/T \sum_{t=0}^{T} W_t((log(H(t)) + 1) - (log(H'(t)) + 1))^2 + 1/T \sum_{t=0}^{T} W_t((log(D(t)) + 1) - (log(D'(t)) + 1))^2$$

**Figure s1.** Multilayer perceptron Neural Network SIR-HCD model architecture.

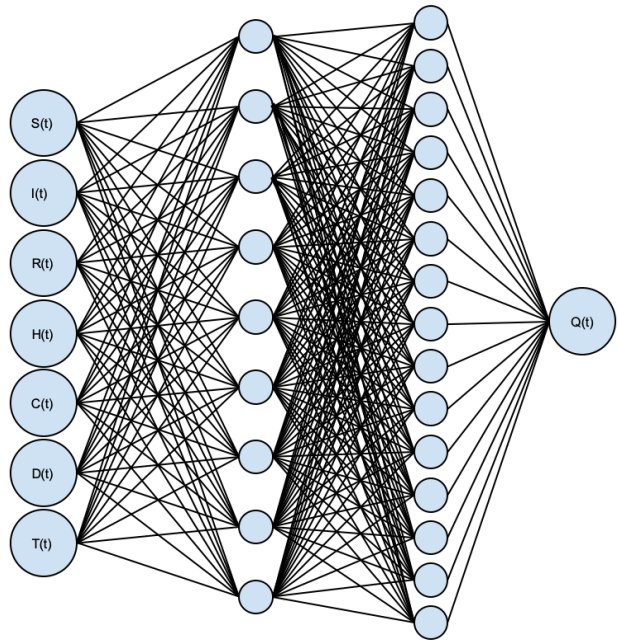

The Neural Network SIR-HCD model does not introduce mixture groups to control idealized policy impact like the SSIR-HCD model. As a result, the Neural Network SIR-HCD model does not perform so well as the SSIR-HCD model (See Figure s2), with higher MLSE in the steep increase at the last time segment of hospitalization data. The quarantine strength term $Q(t)$ is adjustable on a rather small scale; therefore, it fits well on cumulative mortality data. The limitation may originate from the simple design of its network architecture. Applying substantial structure change may allow improvements.

**Figure s2**. Modeled cumulative mortality and hospitalization cases (training + test) in Neural Network SIR-HCD model

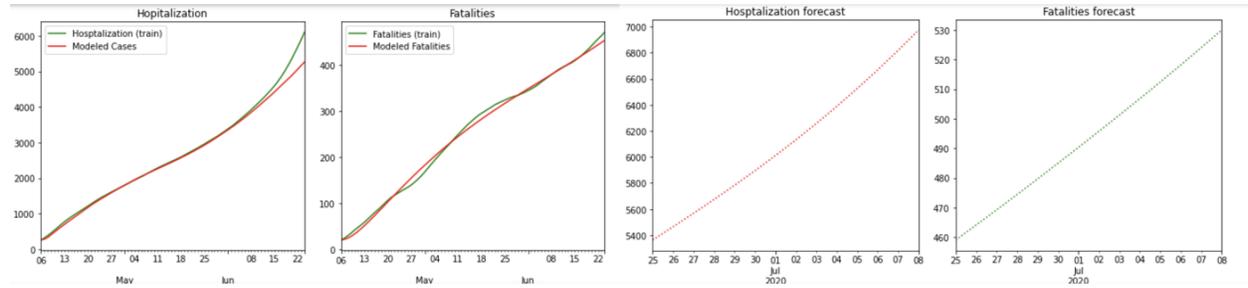

**Figure s3**. Modeled daily hospitalization cases and cumulative mortality (training + test) in Neural Network SIR-HCD model

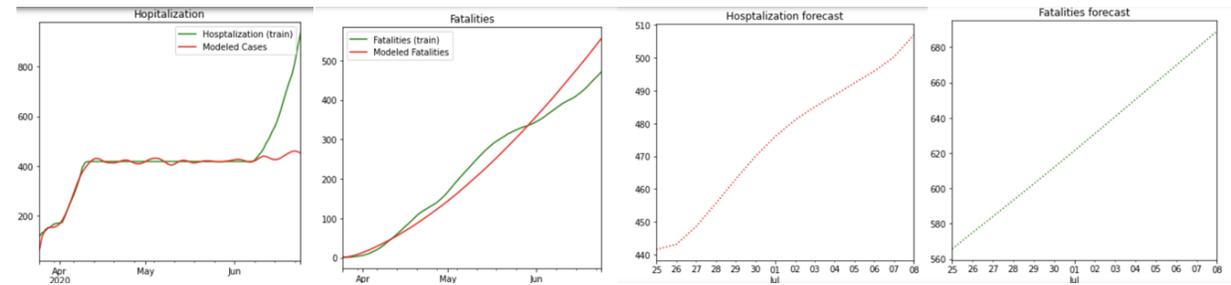